 \documentclass[preprint2]{aastex}
\slugcomment{To appear in ApJ}
\shorttitle{Cloud Core Evolution}
\shortauthors{Steinacker et al.}

\begin{document}

\title{3D Continuum Radiative Transfer Images of a Molecular Cloud Core Evolution}

\author{J. Steinacker}
\affil{Max-Planck-Institut f\"ur Astronomie,
             K\"onigstuhl 17, D-69117 Heidelberg }
\email{stein@mpia.de}
\and

\author{B. Lang}
\affil{Max-Planck-Institut f\"ur Astronomie,
             K\"onigstuhl 17, D-69117 Heidelberg }
\email{lang@mpia.de}

\and

\author{A. Burkert}
\affil{Universit\"ats-Sternwarte M\"unchen,
             Scheinerstr.1, D-81679 M\"unchen }
\email{burkert@usm.uni-muenchen.de}

\and
\author{A. Bacmann}
\affil{Observatoire de Bordeaux, 
             2 Rue de l'Observatoire, BP 89, 33270 Floirac}
\email{bacmann@obs.u-bordeaux1.fr}

\and
\author{Th. Henning}
\affil{Max-Planck-Institut f\"ur Astronomie,
             K\"onigstuhl 17, D-69117 Heidelberg }
\email{henning@mpia.de}

\begin{abstract}
We analyze a 3D Smoothed Particle Hydrodynamics simulation 
of an evolving and later collapsing pre-stellar core.
Using a 3D
Continuum Radiative Transfer program, we generate images at 7, 15, 175 $\mu$m,
and 1.3 mm for different evolutionary times and viewing angles.
We discuss the observability of the properties
of pre-stellar cores for the different wavelengths.
For examples of non-symmetric
fragments, it is shown
that, misleadingly, the density profiles derived from a 1D analysis of
the corresponding images are consistent with 1D core evolution models.
We conclude that 1D modeling based on column density 
interpretation of images does not produce reliable structural
information and that multi-dimensional modeling is
required.
\end{abstract}


% Should have about 100 lines
\section{Introduction}
Molecular cloud cores are thought to be the direct progenitors of
stars. However,
their initial properties and early evolution
are still poorly understood.
Current observations therefore aim to find and study these cores in more 
detail. Density structure, velocity field, and temperature distribution 
of the gas and dust are key parameters for the physical interpretation 
of the long lifetimes of these objects. 
The continuum radiation
spectra of deeply embedded sources contain only ambiguous
information about the density and temperature distribution of the dust.
Column densities can be inferred from the analysis of continuum images in the
mm-range \citep[e.g.][]{War94} and the mid-infrared (MIR) \citep[e.g.][]{Bac00}. 
The emission of molecules within the core contains information about the 
structure,
velocity field \citep[][and references therein]{Taf04}, and the turbulence \citep[][and references therein]{Oss01}.
The currently discussed supporting mechanisms against gravitational collapse
are magnetic
fields and turbulence.
There are a number of simulations treating the full 3D structure of collapsing
pre-stellar cores \citep[e.g.][]{BBB03, BKV03, BurBod93, BurBod00, 
KruFisKle03, KleHeiMac00, HeiMacKle01}. In some of these papers, the
resulting column density along a line of sight was compared
with observationally obtained
column densities.
The question which structures of the distribution can be seen at which
wavelength, however, can only be answered by producing images of the core
for given dust properties.
In turn, the majority of models that have been applied
to observational data are mostly based on spherical symmetry \citep[e.g.][]{And96}.
They rely on spherically symmetric models of isolated star formation, which
describe core formation, gravitational collapse, and proto-stellar accretion.
Commonly, averaged radial density profiles are derived and compared with 
power-law density distributions 
\citep[e.g.][]{And96}, or Bonnor-Ebert spheres
\citep[e.g.][]{Alv04}.
The major difficulties using 3D models in observations are 
i) the information loss due to the projection effect, 
ii) the complex and unique structure of each individual pre-stellar cores, 
and iii) the numerical effort of multi-dimensional radiative transfer.

In this letter, we investigate how an evolving cloud core simulated by
a 3D Smoothed Particle Hydrodynamics (SPH) 
code would appear at different wavelengths, which structures are visible,
and which density 
distributions would be inferred using common 1D models. 
The results from the SPH simulation are described in Sect.~2, along with
a discussion of underlying assumptions.
We present the images from the 3D Continuum Radiative Transfer (CRT) 
modeling and
discuss the observability of different structures and physical effects.
The comparison to density structures obtained from applying 1D models
to the images is given in Sect.~3,
and the findings are summarized and discussed in Sect.~4.

\section{Cloud core evolution model and radiative transfer modeling}
% about 100 lines
\subsection{3D SPH simulation of the evolution of a cloud core}
We have 
calculated the evolution of a cloud core using a three-dimensional
SPH code \citep[version described in][]{BatBonPri95},
originally developed by
\citet{BenCamPre90}.
The smoothing lengths of particles are variable in
time and space, following the constraint that the number
of neighbors for each particle has to be approximately
constant with $N_{\rm neigh}=50$.  The SPH equations are
integrated using a second-order Runge-Kutta-Fehlberg
integrator with individual time steps for each particle
\citep{BatBonPri95}. 

The simulation was initiated with a mass of
$M=3\,\rm{M_{\sun}}$, adopting a spherically
symmetric non-rotating homogeneous cloud with a
temperature of $T=10$ K, a diameter of $d=0.12$ pc (corresponding to
a density of $2\times 10^{-16}$ kg/m$^3$), and a mean molecular
weight $\mu = 2.36\times 10^{-3}$ kg/mol. 
This configuration is Jeans unstable.
A turbulent velocity field is added only at the beginning of
the simulation with a Mach number of ${\cal M}=2$ 
and following an approximate Kolmogorov law
$P(k)d\Omega_k\sim k^{-2}$ for the different modes. The turbulence
supports the cloud core against
collapse for the first $10^5$ yrs.
This enables the formation of a pre-stellar core-like
structure, self-consistently as a result of turbulent energy dissipation.
A variable equation of state is used: isothermal for densities less
than $10^{16}$ molecules/$m^3$ and adiabatic for larger
densities.

\begin{figure}[t]
\vspace{13cm}
\includegraphics{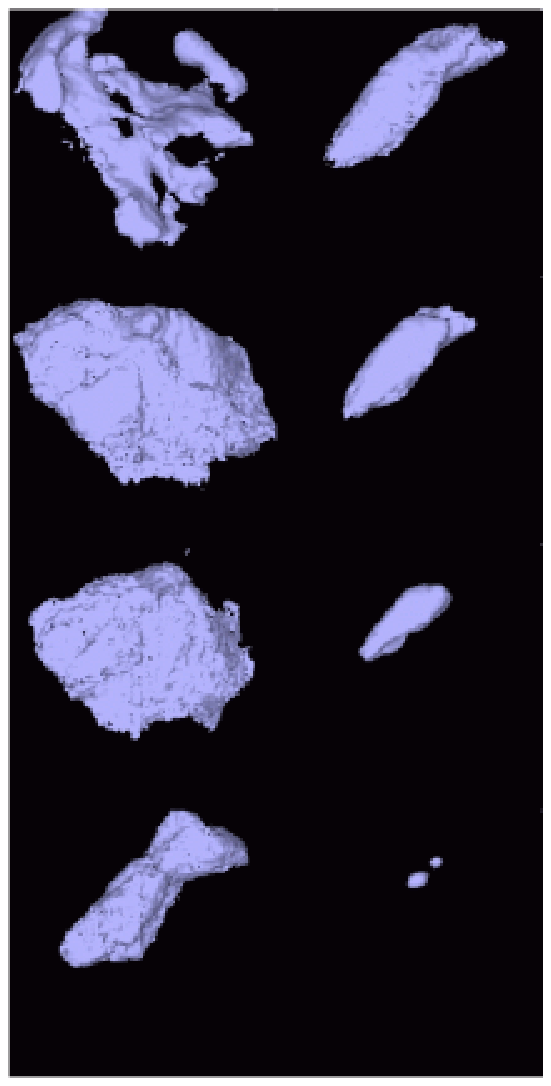}
\caption{Iso-density surfaces of the averaged 
         SPH density distributions of a cloud core fragment.
         The left panel shows densities of 40, 5, 5, and 5 
         $\times 10^{-17}$ kg/m$^3$ at the times
         5.6, 14.1, 16.9, and 27.2$\times 10^4$ yrs
         after the start of simulation, respectively.
         The right panel gives iso-density surfaces at the time 24.4$\times
         10^4$ yrs
         and densities of 0.16, 0.5, 1.6, and 5.2$\times 10^{-18}$ kg/m$^3$,
         respectively.
\label{fig1} }
\end{figure}

Deviating from earlier work,
the initial conditions are arranged in a way that the core reaches a
dynamical equilibrium of density structures and velocity field before
it evolves into a runaway collapse.
The resulting structure is visualized by iso-density surfaces shown
in the left panels of Fig.~\ref{fig1}.
The top left panel shows the early stage of core formation $5.6\times
10^4$ yrs after the 
initialization (iso-density of $4\times 10^{-16}$ kg/m$^3$). 
Turbulence dominates the structure formation and creates several filamentary
low-mass density maxima. 
The duration of this period before the onset of the collapse and thus the total
"age" of the pre-stellar core stage depends on how
turbulence is injected initially and its dissipation.
In the course of time, the local density enhancements merge. 
After some
additional $8.5\times 10^4$ yrs just at the edge of gravitational instability,
a single core has formed (second left panel, $5\times 10^{-17}$ kg/m$^3$). 
The kinetic
pressure support breaks down due to rapid dissipation of turbulent 
energy inside the over-dense region and the core starts to collapse 
($t=1.69 \times10^5$ yrs, third left panel, $5\times 10^{-17}$ kg/m$^3$). 
A new single hydrostatic core forms when
the gas becomes optically thick and the cooling time exceeds the dynamical time.
Lateron, the central part of the core is replaced by a sink particle.
In the bottom left panel, the structure has flattened substantially, 
20\% percent of the total mass is already accreted onto the sink
particle, and a massive disk 
has formed through an instability. 
It contains additional low mass condensations
and independently, a second fragment has started to form with
a hydrostatic core ($5\times 10^{-17}$ kg/m$^3$).   
The right panels give the iso-densities for 0.16, 0.5, 1.6, and 
5.2$\times 10^{-18}$ kg/m$^3$ for a time of $2.4\times 10^5$ yrs, 
respectively. With increasing density, the second 
condensation becomes visible (see also animation at
http://www.mpia-hd.mpg.de/homes/stein/Ani/animcf.htm). 

\subsection{3D CRT modeling of the 
cores}
\begin{figure}[t]
\vspace{15cm}
\includegraphics{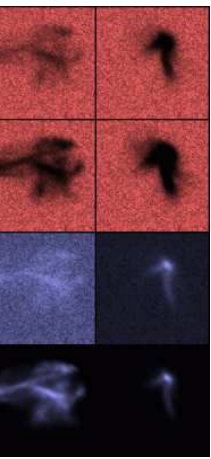}
\caption{Images of the cloud core fragment at the
         wavelengths 7, 15, 175, and 1300 $\mu$m 
         (left columns from top to bottom), and at the
         times 5.6 and 14.1$\times 10^4$ yrs 
         after start of simulation (left to right), respectively.
\label{fig2}}
\end{figure}

The SPH density distributions of the gas were discretized on a 3D grid and
scaled to dust particle distributions assuming a dust-gas mass ratio of 1/100
and an efficient gas-dust mixing.
The dust number densities
were processed with a 3D CRT code
\citep{Ste03, Ste02a, Ste02b, Pas04}, 
producing 640 images of the cloud core at different
wavelengths, times, and viewing angles, respectively. 
The temperatures were calculated from the radiative heating.
Heating by compression is irrelevant during the pre-stellar core phase
due to the fact that the cooling timescale is much faster than the
dynamical timescale. 
For the illustrative purpose of this letter, we used standard dust
opacity 
data \citep{Dra84} and a standard interstellar radiation field
\citep{Bla94}.
Some of the images are shown in Fig.~\ref{fig2} for the wavelengths
7, 15, 175, and 1300 $\mu$m (from top to bottom) and
evolutionary times of 5.6 and 14.1$\times 10^4$ yrs, respectively 
(left to right).  
The wavelengths are chosen to cover common observational windows (e.g. ISOCAM, ISOPHOT,
IRAM, JCMT, CSO, SPITZER, HERSCHEL).
All images are scaled to have
maximal contrast. A 10\% random background noise representing a mean
background variation 
was added for illustrative purpose only.
In the MIR, as expected, the core is visible in absorption and the images reveal
much of the outer thin structure especially for the early stages.
Detection of the inner, at later stages flattened  structure
is difficult and
requires a careful background analysis.
For wavelengths larger than 90 $\mu$m, the cold dust can be seen in emission, 
revealing
more of the inner structure at high densities, as the core also starts to 
get
optically thin. This emission is dominating the mm images.

Animations showing the images for all viewing angles at different times
of the evolution, as well as 
visualizations of the
3D density data cube can be found under
http://www.mpia-hd.mpg.de/homes/stein/Ani/animcf.htm.

\section{1D analysis of the maps}
\begin{figure}[t]
\vspace{1cm}
\includegraphics{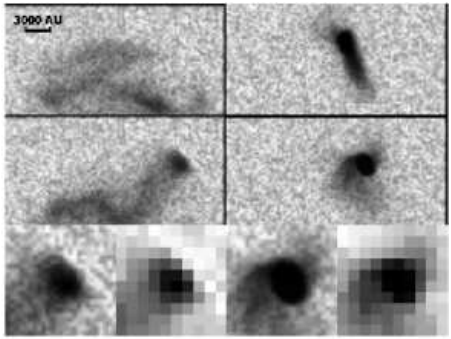}
\caption{Examples of features in 7 $\mu$m images mimicking dense cores.
         The top panels show structures which look like a
         core when choosing
         an appropriate viewing angle (middle panels). The lower panels
         zoom into the core-like structures and switch to ISOCAM resolution.
\label{fig3}}
\end{figure}

\begin{figure}[t]
\vspace{8cm}
\includegraphics{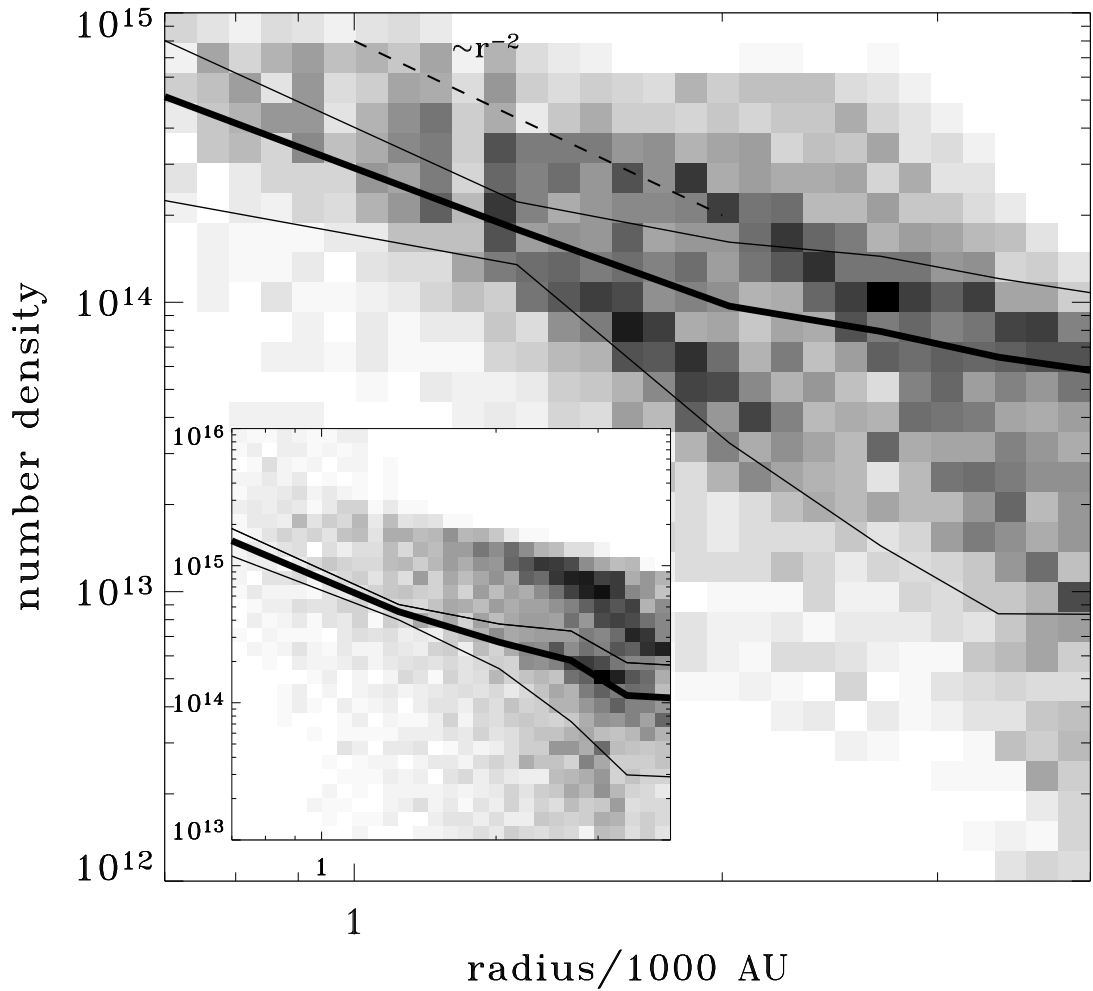}
\caption{1D density profiles $n(r)$ obtained from the 7 $\mu$m-image by
azimuthally integrating the column density 
around the absorption maximum along elliptical tracks.
The thin solid lines mark the range of profiles for individual directions,
and the direction-averaged profile is represented by the thick solid line.
The dashed line
indicates an $r^{-2}$-dependency. 
The true 3D density distribution $n_{SPH}$ was discretized on a cell grid,
and transformed to a point distribution in the $n(r)$-plane. The number
of points is
shown as gray-scale image where black means maximum number of grids
with a certain density.
The main picture corresponds to an elongated filament from the
early stage, and the inlet to a flattened structure from a 
later stage of the core evolution, respectively.
\label{fig4}}
\end{figure}

Projection effects are a severe source of misinterpretation for structures
seen in absorption or emission, as pointed out already, e.g.,
by \citet{BM02}. In Fig.~\ref{fig3}, we show as an example
two structures seen at 7 $\mu$m.
The upper left panel depicts an elongated filament at an early stage of the
evolution ($5.6\times10^4$ yrs), while the upper right panel gives a 
flattened structure at a later stage ($24.4\times10^4$ yrs).
In the middle panels, the
viewing angle was changed until we see the structures as a core-like
feature. In the lower panel, 
they are zoomed and re-binned to an ISOCAM resolution typical for a core at
150 pc distance.
We determined the 1D number column density $N(R)$ with the radius in the
plane of the sky $R$ by azimuthally
averaging over annuli. 
As the absorption patterns have elliptical shape, 
we have used elliptical annuli. The resulting number column density was 
inverted to a 1D number density distribution $n(r)$ with the radius $r$
using recursive integration.
In Fig.~\ref{fig4}, we show the results for the early stage-filament
in the main
panel and for the later stage-disk in the inlet. 
The range of profiles $n(r)$ which have been transformed from
$N(R)$-profiles along individual directions is given by
the solid thin lines and the direction-averaged profile is plotted
as solid thick line.
The dashed line indicates the slope of a density distribution following
a $r^{-2}$-power-law as it was derived from 1D core evolution models 
\citep{Shu77}.
Although these absorption maxima are slightly
less extended than commonly observed cores, the 1D model seems to
provide a reasonable description of the derived distributions.
It could be inferred from this fit that the underlying density structure
has an elliptical shape with a profile that - transformed to a spherical 
distribution - is in agreement with 1D core evolution models.
To compare with the true underlying 3D density distribution,
we calculated the
number density $n$ for a grid of equally-sized cells from the SPH
density distribution. For each cell, we determined the distance to
the center of the "core"-distribution defining a point in the
$n(r)$-diagram. This point distribution was rebinned to a greyscale image
for better clarity, where black refers to maximum number of cells per
bin. The advantage of this representation is that a 1D core with radial
power law appears as a line in the $n(r)$-diagram with a gradient 
representing the powerlaw index.

The true density distributions overlayed as grayscale-image are far from 
being lines as the filament is not 1D spherically symmetric. 
The agreement with the 1D evolution models would tend to validate
static core formation models, although the core formation mechanism
modeled here is highly dynamical. This is in agreement with the
findings of \citet{BKV03}.
We also modelled the flattened structure at a later stage. The radial
profile within the disk will be visible in our
grayscale-representation of $n(r)$ as a line with the slope of the
power-law exponent, and indeed the inlet in Fig.~\ref{fig4}
shows a pronounced branch of the disk-like structure. If the core-flattening
is confirmed by an independent source of information, the column density
profiles (and the resulting density profiles) can be used to determine
the radial profile of the disk-like structure.

% about 40 line
\section{Conclusions}
We have presented 3D simulations assuming
that initially low mass condensations pass through a stage of turbulence dominated condensation where they accumulate mass and merge together to form extended pre-stellar core like objects.
The typical density structures in the cores are non-spherical
throughout their evolution.
The asymmetry is driven by the turbulent motion and causes complex structures
from the very beginning.
This complexity is partially seen in images that have been 
calculated from the densities obtained in the cloud core simulation.
However, projection effects can lead to a severe misinterpretation of 
images.
We showed that
a 1D analysis of the
vicinity of the density maxima would suggest a
density profiles in agreement with 1D core collapse models. 
The underlying density structure, however, is 
intrinsically three-dimensional and deviates strongly from the obtained
1D model distribution.

As the column density also enters the optical thin emission in the
mm-range (aside from the Planck function),
we expect the same projection ambiguities to 
occur when interpretating mm-maps of dense molecular cloud regions. This
aspect will be discussed in a forth-coming paper.

We conclude that 1D modeling based on column density 
interpretation of images does not produce reliable structural
information.
For flattened structures appearing in later stages of the core evolution,
a 2D modeling might be applicable, but for the general case,
multi-dimensional continuum and line radiative
transfer modeling is required to
derive consistent density and temperature distributions
of the gas and dust in pre-stellar
cores.

\clearpage


\clearpage

\clearpage

\clearpage
 
\end{document}